\documentclass[12pt]{article}
\usepackage[dvips]{color}                 
\usepackage{graphicx}                     
\usepackage{amssymb}
\usepackage{amsmath}
\usepackage{xspace}                       
\renewcommand{\today}{\ifcase\day\or 1st\or 2nd\or 3rd\or 4th\or 5th\or 6th\or
7th\or 8th\or 9th\or 10th\or 11th\or 12th\or 13th\or 14th\or 15th\or 
16th\or 17th\or 18th\or 19th\or 20th\or 21st\or 22nd\or 23rd\or 24th\or
25th\or 26th\or 27th\or 28th\or 29th\or 30th\or 
31st\fi~\ifcase\month\or January\or February\or March\or April\or
May\or June\or July\or August\or September\or October\or November\or
December\fi \space \number\year}   
\newcommand{\mytitle}[1]{
\begin{center}
\LARGE{\textbf{#1}}
\end{center}}
\newcommand{\myauthor}[1]{\textbf{#1}}
\newcommand{\myaddress}[1]{\textit{#1}}
\newcommand{\mypreprint}[1]{\begin{flushright} #1 \end{flushright}}
\begin{document}
%
\begin{titlepage}
\mypreprint{
\textbf {TUM-T39-03-17} \\
}
\vspace*{1.0cm}

\mytitle{Nucleon mass, sigma term and lattice QCD\footnote{Work supported in part by BMBF and DFG.}}
\vspace*{1cm}
\begin{center}
 \myauthor{Massimiliano Procura$^{a,b}$}, \myauthor{Thomas R. Hemmert$^{a}$}
and
\myauthor{Wolfram Weise$^{a,b}$} 
\vspace*{0.5cm}

\myaddress{$^a$ Physik-Department, Theoretische Physik  \\
    Technische Universit{\"a}t M{\"u}nchen, D-85747 Garching, Germany\\
    (Email: themmert@physik.tu-muenchen.de)}\\[2ex]
\myaddress{$^b$ ECT*, Villa Tambosi, I-38050 Villazzano (Trento), Italy\\
    (Email: weise@ect.it, procura@ect.it)}
\vspace*{0.2cm}
\end{center}

\vspace*{2.5cm}

\begin{abstract}
We investigate the quark mass dependence of the nucleon mass $M_N$. An interpolation of this observable, between a selected set of fully dynamical two-flavor lattice QCD data and its physical value, is studied using relativistic baryon chiral perturbation theory up to order $p^4$. In order to minimize uncertainties due to lattice discretization and finite volume effects our numerical analysis takes into account only simulations performed with lattice spacings $a<0.15\,{\rm{fm}}$ and $m_\pi\,L>5$. We have also restricted ourselves to data with $m_\pi<600\,{\rm{MeV}}$ and $m_{\rm{sea}}=m_{\rm{val}}$. A good interpolation function is found already at one-loop level and chiral order $p^3$. We show that the next-to-leading one-loop corrections are small. From the $p^4$ numerical analysis we deduce the nucleon mass in the chiral limit, $M_0 \approx 0.88\,{\rm{GeV}}$, and the pion-nucleon sigma term $\sigma_N= (49 \pm 3)\,{\rm{MeV}}$ at the physical value of the pion mass. 
\end{abstract}


\noindent

\end{titlepage}
\setcounter{page}{2} 
\newpage


\section{Introduction and framework}

Lattice QCD on one side and chiral effective field theory, on the other, are progressively developing as important tools to deal with the non-perturbative nature of low-energy QCD and the structure of hadrons \cite{TW}. The merger of both strategies has recently been applied to extract
physical properties of hadrons---such as the nucleon---from lattice QCD simulations. Of particular interest in such extrapolations is the detailed quark mass dependence of nucleon properties. Examples of recent extrapolation studies concern the nucleon mass \cite{Ross, BHM},  its axial vector coupling constant and magnetic moments \cite{AUS,MUN}, form factors \cite{GOEC} and moments of structure functions \cite{Negele}.

Accurate computations of the nucleon mass with dynamical fermions and two active flavors are now possible \cite{CPPACS, JLQCD, QCDSF} in lattice QCD. However, the masses of $u-$ and $d-\,$quarks used in these evaluations exceed their commonly accepted small physical values, typically by an order of magnitude. It is at this point where chiral effective field theory methods are useful - within limitations discussed extensively in refs. \cite{Ross, BHM} - in order to interpolate between lattice results, actual observables and the chiral limit ($m_{u,d} \to 0$). In this paper we explore the capability of such an approach for extracting the nucleon mass and the pion-nucleon sigma term.

The nucleon mass is determined by the expectation value $\langle N\,|\Theta_\mu^{\mu}|\,N \rangle$ of the trace of the QCD energy-momentum tensor \cite{DGH},
\begin{eqnarray}
\Theta_\mu^{\mu}&=&\frac{\beta(g)}{2g}G_{\mu \nu}G^{\mu \nu} + m_u \bar{u}u+m_d \bar{d}d+... ~~  ,
\end{eqnarray}
where $G^{\mu \nu}$ is the gluonic field strength tensor, $\beta(g)$ is the beta function of QCD and $m_q \bar{q}q$ with $q=u,\,d \dots$ are the quark mass terms (we omit here the anomalous dimension of the mass operator, as in \cite{Ji}).  
So the physical nucleon mass $M_N$ can be expressed as
\begin{eqnarray}
M_N&=& M_0+\sigma_N
\end{eqnarray}
in terms of its value $M_0$ in the $SU(2)_f$ chiral limit,
\begin{eqnarray}
M_0&=&\langle N\,|\frac{\beta}{2g} G_{\mu \nu}G^{\mu \nu}+ \dots|\,N \rangle 
\end{eqnarray}
(with suitably normalized nucleon Dirac spinors). The dots refer to possible contributions from heavier quarks, other than $u$ and $d$, and the sigma term is defined as
\begin{eqnarray}
\sigma_N&=&\! \sum_{q=u,d}  m_q \frac{{\rm{d}} M_N}{{\rm{d}}m_q}\,=\,\langle N\,|m_u \bar{u}u+m_d \bar{d}d|\,N \rangle~~~.
\label{sigma}
\end{eqnarray}
The quark mass dependence of $M_N$ translates into a dependence on the pion mass: $m_\pi^2 \sim m_q$ at leading order. We pursue this connection in the symmetry breaking part of the chiral effective Lagrangian. 

The framework of our study is relativistic $SU(2)_f$ baryon chiral perturbation theory (BChPT) as described in ref.\cite{BL}. The effective Lagrangian required for our analysis of the nucleon mass up to ${\cal{O}}(p^4)$ is 
\begin{eqnarray}
{\cal{L}}&=&{\cal{L}}^{(1)}_N+{\cal{L}}^{(2)}_N+{\cal{L}}^{(4)}_N+{\cal{L}}^{(2)}_{\pi} 
\end{eqnarray}
with
\begin{eqnarray}
{\cal{L}}_N^{(1)}&=&\bar{\Psi}\,(i\gamma_\mu D^{\mu}-M_0)\,\Psi+\frac{1}{2}\,g_A\, \bar{\Psi}\,\gamma_\mu \gamma_5 u^{\mu}\, \Psi\,, \nonumber\\
{\cal{L}}_N^{(2)}&=&c_1\,{\rm{Tr}}(\chi_{+}) \bar{\Psi} \Psi-\frac{c_2}{4M_0^2}\,{\rm{Tr}}(u_\mu u_\nu)\,(\bar{\Psi} D^{\mu}D^{\nu} \Psi+ {\rm{h.c.}})+\frac{c_3}{2}\,{\rm{Tr}}(u_\mu u^{\mu})\, \bar{\Psi}\Psi+... \nonumber\\
{\cal{L}}_N^{(4)}&=&e_{38}\,({\rm{Tr}}(\chi_{+}))^2 \bar{\Psi}\Psi+\frac{e_{115}}{4}\,{\rm{Tr}}(\chi_{+}^2-\chi_{-}^2)\bar{\Psi}\Psi\nonumber\\&&-\frac{e_{116}} {4}\,\left({\rm{Tr}}(\chi_{-}^2)-({\rm{Tr}}(\chi_{-}))^2+{\rm{Tr}}(\chi_{+}^2)-({\rm{Tr}}(\chi_{+}))^2\right) \bar{\Psi}\Psi +\dots~~~.
\end{eqnarray}
In ${\cal{L}}_N^{(4)}$ we follow the notation of ref.\cite{fettes}.
Here ${\cal{L}}_\pi^{(2)}$ is the leading order pion Lagrangian including the mass term. The nucleon Dirac field is denoted by $\Psi$, and $M_0$ is the nucleon mass in the chiral limit. The axial field
$u^{\mu}$ and the covariant derivative $D^{\mu}$ involve the Goldstone boson fields via $U(x) \in SU(2)$, and $\chi_{\pm}=u^{\dag} \chi u^{\dag} \pm u \chi^{\dag} u$, $u^2=U$, parametrizes the explicit chiral symmetry breaking through the quark masses; here we use $\chi=2B {\cal{M}}$, where ${\cal{M}}={\rm{diag}}(m_u,m_d)$ and $B= -\langle\bar{q}q \rangle/f_\pi^2$ is the chiral condensate divided by the pion decay constant squared, both taken in the chiral limit. In the following we neglect isospin breaking effects.

\section{Analytic results}
\subsection{${\cal O}(p^3)$ Analysis}

\begin{figure}[t]
  \begin{center}
    \includegraphics*[width=0.7\textwidth]{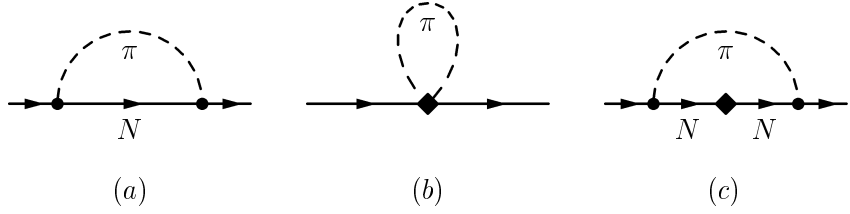}
    \caption{One-loop graphs of NLO (a) and NNLO (b, c) contributing to the nucleon mass shift. The solid dot denotes a vertex from ${\cal{L}}_N^{(1)}$, the diamond a vertex from ${\cal{L}}_N^{(2)}$.}
    \label{diags}
  \end{center}
\end{figure}

The leading order contribution to the shift of the nucleon mass from its value in the chiral limit comes from the explicit chiral symmetry breaking piece in ${\cal{L}}_N^{(2)}$, which drives the nucleon sigma term $\sigma_N$ of Eq.(4). The next-to-leading order (NLO) contribution is represented by diagram (a) of Fig.\ref{diags}, with the $\pi N N$ vertex generated by ${\cal{L}}_N^{(1)}$. We have evaluated the relevant one-loop integrals using the so-called infrared regularization method \cite{BL}. It represents a variant of dimensional regularization which treats one-loop integrals involving baryon propagators in a way consistent with chiral power-counting. The diagram (a) develops a divergence proportional to $m_\pi^4$. It is absorbed in contact terms which are formally of fourth order. We denote\footnote{Our coupling $e_1$ differs from the convention of ref.\cite{BHM} by a factor 4.} the counterterm structure that renders the ${\cal O}(p^3)$ contribution finite as $-e_1 m_\pi^4 \bar{\Psi}\Psi$ . In the notation of ref.\cite{fettes} it involves the coupling constant combination $e_1=-(16e_{38}+2e_{115}+2e_{116})$ from ${\cal{L}}_N^{(4)}$.

Following the reasoning outlined here, the constraint to obtain a finite result at leading one-loop order has effectively promoted a linear combination of $p^4$ couplings---denoted by $e_1$---into the $p^3$ calculation. The resulting expression for the $m_\pi$ dependence of $M_N$ then reads:
\begin{eqnarray}
M_N&=&M_0-4c_1 m_\pi^2+ \left[e_1^r(\lambda)+\frac{3{g_A}^2}{64 \pi^2 f_\pi^2	M_0 }(1-2 \ln{\frac{m_\pi}{\lambda}})\right]m_\pi^4\nonumber\\&& -\frac{3{g_A}^2}{16 \pi^2 f_\pi^2}\, m_\pi^3\, \sqrt{1-\frac{m_\pi^2}{4 M_0^2}}\left[\frac{\pi}{2}+\arctan{\frac{m_\pi^2}{\sqrt{4 M_0^2 m_\pi^2- m_\pi^4}}}\right]. \label{massa} 
\end{eqnarray} 
Here $e_1^r(\lambda)$ is the finite (renormalization scale $\lambda$ dependent) part of $e_1$,
\begin{eqnarray}
e_1&=&e_1^r(\lambda)+\frac{3\,{g_A}^2}{2f_\pi^2\, M_0}\,L ~~~, \nonumber
\end{eqnarray}
and any ultraviolet divergences appearing in the limit $d \to 4$ are subsumed in
\begin{eqnarray}
L&=&\frac{{\lambda}^{d-4}}{16 \pi^2}\left[\frac{1}{d-4}-\frac{1}{2}(\ln{(4\pi)}+\Gamma'(1)+1)\right]. \nonumber
\end{eqnarray}
For further discussion we expand the ${\cal O}(p^3)$ result Eq.(\ref{massa}) in powers of the pion mass and obtain
\begin{eqnarray}
M_N&=&M_0-4c_1 m_\pi^2-\frac{3{g_A}^2}{32 \pi f_\pi^2 }m_\pi^3\nonumber\\&&+\left[e_1^r(\lambda)-\frac{3{g_A}^2}{64 \pi^2 f_\pi^2  M_0}(1+2\ln{\frac{m_\pi}{\lambda}})\right]m_\pi^4\nonumber\\&&+\frac{3{g_A}^2}{256 \pi f_\pi^2  M_0^2}m_\pi^5+ {\cal{O}}(m_\pi^6). \label{massachir} 
\end{eqnarray} 
Note that the sum of the first three terms in this formula coincides with the well-known leading one-loop expression for $M_N$ of Heavy Baryon Chiral Perturbation Theory (HBChPT), as expected in the infrared regularization approach \cite{BL}.
From Eq.(\ref{massachir}) one can also deduce that the counterterm $e_1$ of Eq.(\ref{massa}), required in relativistic baryon ChPT for renormalization purposes, is equivalent to the counterterm introduced in \cite{BHM} which regularizes the short distance behaviour in HBChPT. 

In ref.\cite{BHM} an assessment of convergence properties of ChPT has been performed. It was shown that, up to pion masses around $m_\pi \simeq 300$ MeV, the chiral perturbation expansion develops a stable plateau region independent of cutoff scales. Moreover, this analysis indicates that an upper limit for this plateau behaviour may be reached when $m_\pi$ approaches about 600 MeV. While these considerations were made in a non-relativistic framework, explicit comparison shows that the relativistic approach used in the present work contains the same chiral structures as those discussed in ref.[3]. We can therefore assume that, with respect to the internal consistency of ChEFT, our analysis is applicable for pion masses well above the physical one. We give $m_\pi \lesssim 600\,{\rm{MeV}}$ as an estimate for the range of validity. We emphasize that, in contrast to the framework adopted by the Adelaide group \cite{Ross}, {\it all the terms} beyond the leading $c_1$ contribution to the nucleon mass in Eq.(\ref{massachir}) are part of the {\it same} chiral order $p^3$. The numerical evaluation of the individual contributions to Eq.(\ref{massachir}) (see Section 3) shows that the large fluctuations in the chiral extrapolation using dimensional regularization, reported in \cite{Ross}, arise from examining only the first four terms in Eq.(\ref{massachir}), instead of keeping the full expression (\ref{massa}).

\subsection{${\cal O}(p^4)$ Analysis}

Let us now focus on the next-to-NLO (NNLO) contribution to the pion mass dependence of the nucleon mass. This involves also graphs (b) and (c) in Fig.\ref{diags} which include vertices generated by  ${\cal{L}}_N^{(2)}$ as well as wave-function renormalization \cite{BL}. In order to avoid having to deal with a number of counterterms too large to be handled in a meaningful numerical analysis, we decide to truncate the chirally expanded formula at ${\cal{O}}(m_\pi^6)$. We will show numerically that this truncation approximates the full function reasonably well for the parameter ranges considered here. Up to terms of order $m_\pi^6$ no counterterms other than $e_1$ are required for a finite result. At ${\cal O}(p^4)$ one then obtains
\begin{eqnarray}
M_N&=&M_0-4c_1 m_\pi^2-\frac{3{g_A}^2}{32 \pi f_\pi^2 }m_\pi^3\nonumber\\
&+& \left[e_1^r(\lambda)-\frac{3}{64\pi^2 f_\pi^2}\left(\frac{{g_A}^2}{M_0}-\frac{c_2}{2}\right) -\frac{3}{32 \pi^2 f_\pi^2 }\left(\frac{{g_A}^2}{M_0}-8 c_1 +c_2 +4c_3\right)\ln{\frac{m_\pi}{\lambda}}\right]\,m_\pi^4 \nonumber\\
&+&\frac{3 {g_A}^2}{256 \pi f_\pi^2 M_0^2} m_\pi^5 + {\cal{O}}(m_\pi^6),
\label{massap4}  
\end{eqnarray}
where now
\begin{eqnarray}
e_1&=&e_1^r(\lambda)+\frac{3 L}{2 f_\pi^2}\left(\frac{{g_A}^2}{M_0}-8 c_1+c_2+4 c_3\right). \nonumber
\end{eqnarray}
This expression includes the constants $c_2$ and $c_3$ which encode the influence of  the $\Delta(1232)$ resonance in low energy pion-nucleon scattering. The terms up to $m_\pi^4$ have already been discussed in \cite{GSS}. For related discussions of the $SU(3)_f$ case see \cite{bugra} and references therein. It is also interesting to observe that our truncation of the relativistic result at $m_\pi^6$ as shown in Eq.(\ref{massap4}) formally coincides with the expansion of nucleon mass in HBChPT to fifth order, as there are no genuine two-loop graph contributions at this order in the chiral expansion \cite{birse}.

\section{Numerical analysis and contact with lattice QCD}

We proceed with the numerical evaluation of Eqs.(\ref{massa}) and (\ref{massap4}). We set
the nucleon axial vector coupling and the pion decay constant equal to their physical values, $g_A=1.267$ and $f_\pi=92.4$ MeV. Strictly speaking, these quantities should be taken in the chiral limit. We have checked that using current estimates for $g_A^0$ and $f_\pi^0$ at $m_\pi \rightarrow 0$ does not lead to any significant changes in our final results. Details are discussed in the last part of this section. Without loss of generality we choose $\lambda=1$ GeV. At order $p^3$ we are then left with three unknown parameters ($M_0$, $c_1$ and $e_1^r(1 {\rm GeV})\equiv \hat{e}_1$) and four parameters at order $p^4$ ($M_0$, $c_1$,  $A\equiv e_1^r(1 {\rm GeV})+3 c_2/(128 \pi^2 f_\pi^2)$ and $B\equiv c_2 + 4c_3$). Our NNLO result is identified with Eq.(\ref{massap4}). This limits the number of tunable coefficients but still keeps sufficiently many orders in $m_\pi$ to provide a good approximation to the full ${\cal{O}}(p^4)$ result. 

The unknown parameters are determined using as input a combined set of lattice QCD data obtained by the CP-PACS \cite{CPPACS}, JLQCD \cite{JLQCD} and QCDSF \cite{QCDSF} collaborations. These computations are performed using fully dynamical quarks with two flavors. In order to minimize artifacts from discretization and finite volume effects, we have selected from the whole set of available data those with lattice spacings $a< 0.15\,{\rm{fm}}$ and  $m_\pi\,L>5$. Furthermore we restrict ourselves to the resulting four data points with $m_\pi<600\,{\rm{MeV}}$ and with equal valence and sea quark masses, $m_{\rm{sea}}=m_{\rm{val}}$. A study of finite volume dependence is in preparation \cite{arifa}. We have expressed lattice data in physical units via the Sommer scale $r_0=0.5\,{\rm{fm}}$ \cite{Sommer}, not taking into account systematic errors arising from possible quark mass dependence of $r_0$ occuring in dynamical simulations. A forthcoming study will address this issue \cite{PHW}.

In a preliminary step we have fitted the set of lattice points using the LO result treating $M_0$ and $c_1$ as free parameters. The resulting linear fit gives an estimate of $c_1$ about a factor 3 smaller than the value determined in $\pi N$ scattering analyses with $\chi^2_{\rm{d.o.f.}}=2.25$. We conclude that linear fits in the quark mass are not appropriate to describe the quark mass dependence of baryon masses. 

\begin{figure}[t]
  \begin{center}
    \includegraphics*[width=0.7\textwidth]{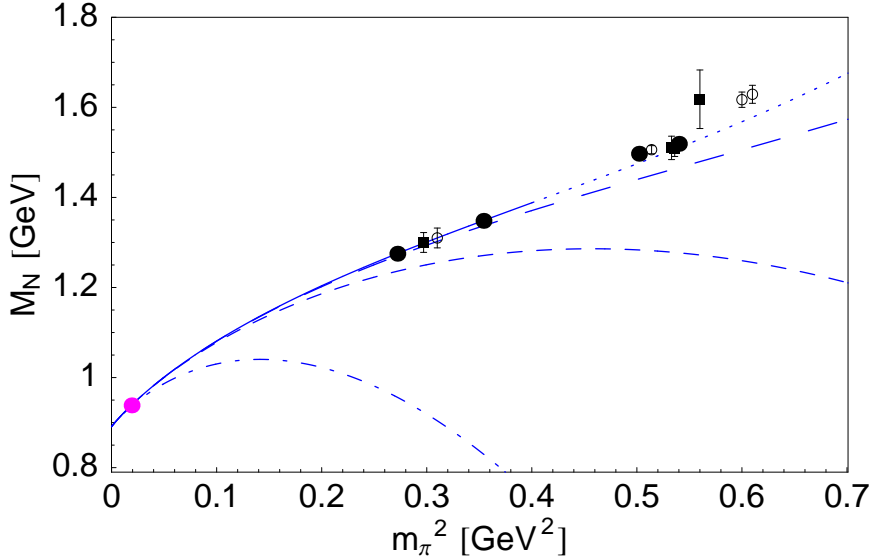}
    \caption{Solid/dotted line: best fit curve using the one-loop result at chiral order $p^3$ Eq.(\ref{massa}). Input: four lowest lattice data points with $m_\pi<600\,{\rm{MeV}}$ and physical nucleon mass (Fit I). The dotted extension of this curve for $m_\pi^2>0.4\,{\rm{GeV}^2}$ indicates the region where the application of baryon ChPT is usually believed to become unreliable. For illustration we also show a subset of lattice data up to $m_\pi \approx 0.8\,{\rm{GeV}}$, those compatible with the cuts in lattice spacing and volume as explained in the text. The solid dots are CP-PACS data, the boxes refer to JLQCD and the empty circles to QCDSF. The dot-dashed, dashed and long-dashed curves show, respectively, the contributions from the sum of the first three, four and five terms in Eq.(\ref{massachir}).}
    \label{figp3}
  \end{center}
\end{figure}

\begin{figure}[t]
  \begin{center}
    \includegraphics*[width=0.7\textwidth]{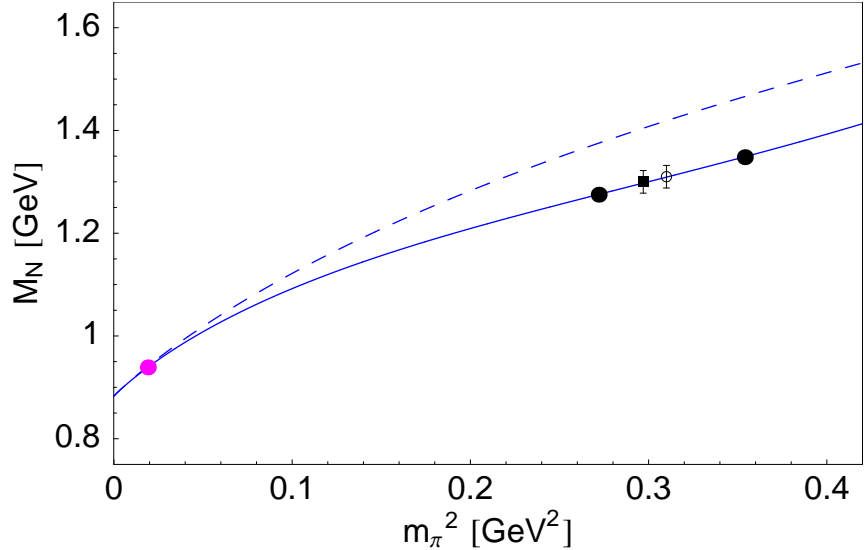}
    \caption{Solid curve: best fit (Fit II) using the NNLO result, Eq.(\ref{massap4}), at chiral order $p^4$. Dashed curve: NLO result (chiral order $p^3$) from Eq.(\ref{massa}) using as parameters the central values of Fit II. $\hat{e}_1$ has been deduced setting $c_2=3.2\,{\rm GeV}^{-1}$. For details on the data points see Fig.\ref{figp3}.}
    \label{figp4}
  \end{center}
\end{figure}

Successive steps in our analysis are shown in Figs.\ref{figp3}, \ref{figp4} and summarized in table \ref{table1}.
We have first analyzed the ${\cal{O}}(p^3)$ result, Eq.(\ref{massa}) (Fit I). The best Fit I curve is the solid one drawn in Fig.\ref{figp3}. The low-energy constants come out of natural size. Furthermore, $c_1$ which determines the slope of $M_N(m_\pi^2)$ for small $m_\pi^2$ has the correct sign and the value of $\hat{e}_1$ is within the range quoted in ref.\cite{BHM}.

As seen in Fig.\ref{figp3}, the curve obtained by fitting the four lattice data with $m_\pi<600\,{\rm{MeV}}$  and including the physical point as a constraint shows a surprisingly good (and not yet understood) agreement with lattice data even up to $m_\pi\approx 750\,{\rm{MeV}}$. 

The same figure also shows how Fit I develops term by term when the full order $p^3$ NLO expression (\ref{massa}) is expanded according to Eq.(\ref{massachir}). We emphasize again that in the hierarchy of terms with increasing powers of $m_\pi$, as represented by the dash-dotted, short-dashed and long-dashed curves, all contributions are of the same chiral order $p^3$ in the formulation of baryon ChPT we use. Evidently, truncating the expansion (\ref{massachir}) at $m_\pi^5$ already provides a decent approximation to the full ${\cal O}(p^3)$ result.

In the NNLO case the statistics of our restricted data sample is not sufficient to constrain all the parameters. We have therefore used input values for $c_2$ and $c_3$ available in the literature. We set $c_2=3.2\,{\rm GeV}^{-1}$ in agreement with refs.\cite{FMS, bernard} and performed two kinds of fits, one with $c_3=-3.4\,{\rm GeV}^{-1}$, found in \cite{EM} to be consistent with empirical $NN$ phase shifts and still within the error bar quoted in \cite{BM} (Fit II), and another one with $c_3=-4.7\,{\rm GeV}^{-1}$, the central value determined by \cite{BM} in low-energy $\pi N$ scattering analysis (Fit III). 

Fit III underestimates $c_1$, whereas the value obtained in Fit II for this LEC is in agreement with \cite{BM} and with the outcome of the analysis by Becher and Leutwyler of low-energy $\pi N$ scattering in $SU(2)_f$ relativistic baryon ChPT \cite{BL2}, the framework we use. Furthermore, the value for $c_3$ employed in Fit II is quite close to $-2.9\,{\rm GeV}^{-1}$, the one corresponding to the empirical spin-isospin averaged p-wave scattering volume which is dominated by the $\Delta(1232)$ contribution. 

Fig.\ref{figp4} demonstrates that the difference between the ${\cal{O}}(p^4)$ and ${\cal{O}}(p^3)$ results is relatively small over the entire range of $m_\pi$ that we analyzed. We explicitly show that higher order chiral corrections are small, even at pion masses well above the physical one. We therefore believe that our interpolation has passed the necessary tests of consistency and convergence for $m_\pi <0.6\,{\rm{GeV}}$.

In the calculations underlying the Fits I-III we have used $g_A=1.267$ and $f_\pi=92.4\,{\rm{MeV}}$ as input. Rigorously speaking, we should have used values of those quantities in the chiral limit. We have performed test calculations with $g_A^0=1.2$ \cite{MUN} and $f_\pi^0=88\,{\rm{MeV}}$ \cite{GL}. With a slight re-adjustment of $\hat{e}_1$ by less than 3\%, any one of the quantities in table \ref{table1} changed by less than 1\% when replacing $g_A$, $f_\pi$ with $g_A^0$, $f_\pi^0$.

\subsection{The sigma term of the nucleon}

The pion-nucleon sigma term $\sigma_N$, defined in Eq.(\ref{sigma}), translates into
\begin{eqnarray}
\sigma_N&=& m_\pi^2 \frac{\partial M_N}{\partial m_\pi^2}~~,
\label{sigma2}
\end{eqnarray}
if we assume that the Gell-Mann - Oakes - Renner  (GOR) relation $m_\pi^2\sim m_q$ holds and we can neglect ${\cal{O}}(m_q^2)$ terms. An improved analysis of s-wave $\pi\pi$ scattering lengths \cite{CGL} indicates that the ${\cal{O}}(m_q^2)$ corrections to the GOR relation are very small, although this statement becomes progressively less accurate with increasing quark masses, and further detailed examination of the role of strange quarks in this context is necessary. A recent systematic analysis \cite{Duerr} of results for pseudo-Goldstone boson masses from $N_f = 2$ lattice QCD comes to conclusions consistent with those drawn in ref.\cite{CGL}. We therefore use Eq.{(\ref{sigma2}) in the following.

At chiral order $p^3$ starting from Eq.(\ref{massachir}) one finds the expression
\begin{eqnarray}
\sigma_N&=& -4c_1 m_\pi^2-\frac{9{g_A}^2}{64 \pi f_\pi^2}m_\pi^3\nonumber\\
&+&2e_1^r(\lambda)m_\pi^4-\frac{3{g_A}^2}{64 \pi^2 f_\pi^2  M_0}(3+4\ln{\frac{m_\pi}{\lambda}})m_\pi^4\nonumber\\&+&\frac{15{g_A}^2}{512 \pi f_\pi^2  M_0^2}m_\pi^5+ {\cal{O}}(m_\pi^6).
\end{eqnarray}
The corresponding NNLO result is derived from Eq.(\ref{massap4}): 
\begin{eqnarray}
\sigma_N &=& -4c_1 m_\pi^2-\frac{9{g_A}^2}{64 \pi f_\pi^2}m_\pi^3\nonumber\\
&+& \left[2e_1^r(\lambda)- \frac{1}{16 \pi^2 f_\pi^2}\left(\frac{9g_A^2}{4M_0} - 6c_1 + 3c_3\right) -\frac{3}{16 \pi^2 f_\pi^2}\left(\frac{g_A^2}{M_0} - 8c_1 + c_2 + 4c_3\right) \ln{\frac{m_\pi}{\lambda}}\right]m_\pi^4\nonumber\\
&+&\frac{15{g_A}^2}{512 \pi f_\pi^2  M_0^2}m_\pi^5+ {\cal{O}}(m_\pi^6). \label{sigmap4}
\end{eqnarray}
Our deduced values of $\sigma_N$ at the physical point are summarized in table \ref{table2}.  The behaviour of the sigma term as a function of the pion mass is shown in Fig.\ref{figsigma}. Within errors, this curve is compatible with the "empirical" sigma term $\sigma_N = 45\pm8$ MeV extracted in ref.\cite{GLS}, but it does not favor the much larger value reported in ref.\cite{PASW}. Our result is also consistent with the analysis of ref.\cite{LTW}, within the larger uncertainties quoted there.

\begin{figure}[t]
 \begin{center}
 \includegraphics*[width=.48\linewidth]{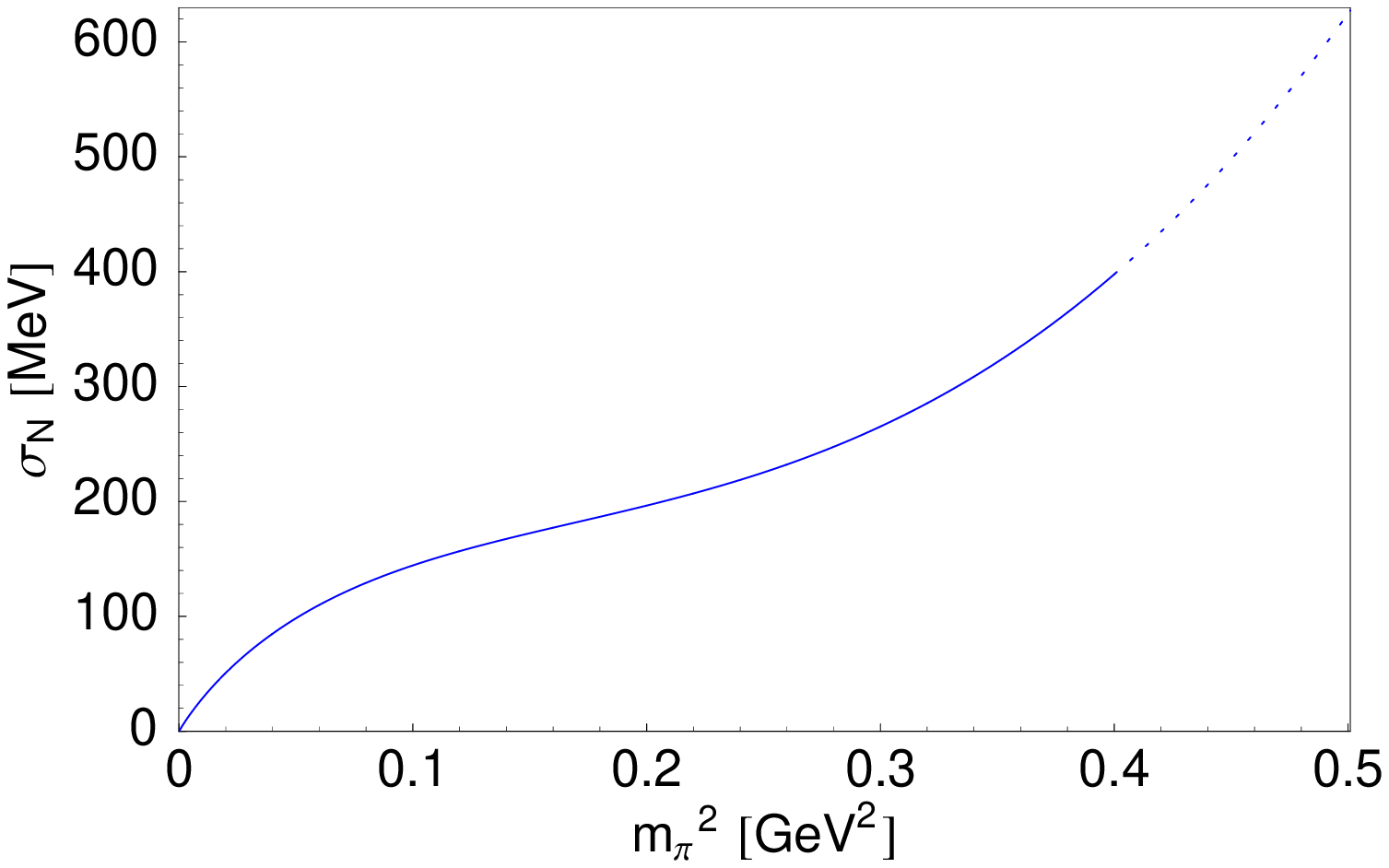}
 \includegraphics*[width=.48\linewidth]{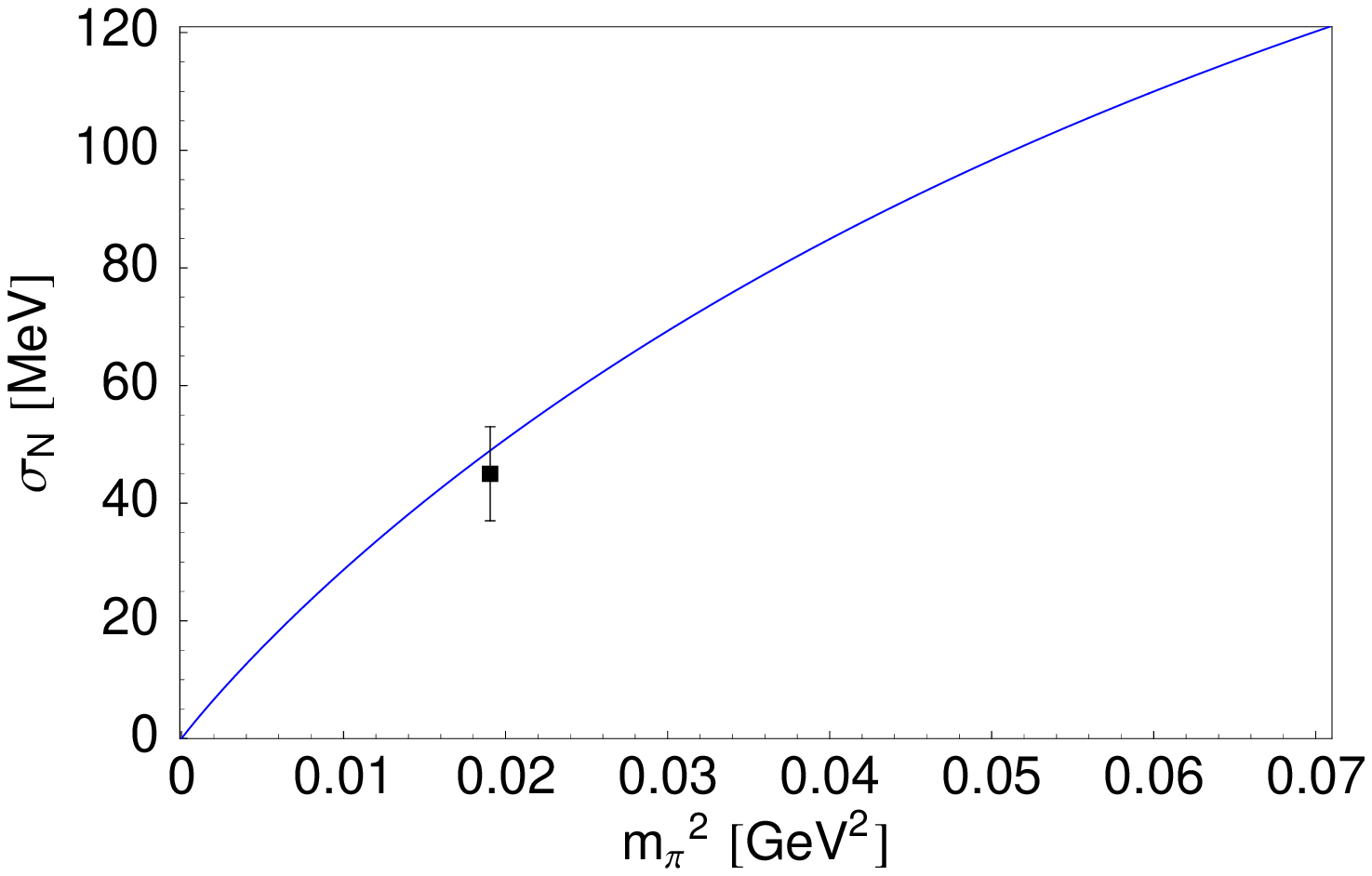}
  \caption{The pion-nucleon sigma term as a function of $m_\pi^2$ from Eq.(\ref{sigmap4}), using as input the central values from Fit II (see table \ref{table1}). The small $m_\pi$ region is magnified in the right panel and plotted together with the frequently quoted empirical $\sigma_N=45 \pm 8\, {\rm{MeV}}$ \cite{GLS}.} \label{figsigma}
 \end{center}
\end{figure}

\section{Discussion and conclusions}

The present work has been aimed at improving and updating interpolations of the nucleon mass, using chiral effective field theory, between the range of relatively large quark masses accessible in full lattice QCD simulations, and the small quark masses relevant for comparison with physical observables. A remarkably good interpolation can already be achieved by a one-loop calculation at chiral order $p^3$ using relativistic baryon ChPT. In either case short distance dynamics, including effects of the $\Delta(1230)$ and possibly other resonance excitations of the nucleon, are encoded in a single counterterm that controls the contributions of order $m_\pi^4$. ${\cal{O}}(p^3)$ relativistic baryon ChPT is therefore free of limitations of HBChPT discussed in \cite{Ross,BHM}. Apart from the nucleon mass in the chiral limit, the only remaining parameter, $c_1$, drives the pion-nucleon sigma term. Our interpolation is based on a selected set of lattice data corresponding to the largest available lattice volumes and the lowest available pion masses, in order to minimize uncertainties from finite size effects and from quark masses too large to be handled using perturbative chiral expansions. Surprisingly, the resulting interpolations work even in a pion mass region where the approach is commonly believed to become unreliable.

The extension to NNLO (chiral order $p^4$), truncated at order $m_\pi^5$, introduces in addition the pion-nucleon low-energy constants $c_{2,3}$ which primarily reflect the impact of $\Delta$ resonance physics on low-energy $\pi N$ dynamics. We have constrained the input values for these two LECs from $\pi N$ phenomenology. The fit interpolating the nucleon mass between the chiral limit and the lattice data remains remarkably stable and even improves slightly when going from NLO to NNLO. Our analysis explicitly shows that ${\cal{O}}(p^4)$ corrections are small with respect to the ${\cal{O}}(p^3)$ result.

The pion-nucleon sigma term deduced from the $m_\pi$ dependence of the nucleon mass in the NNLO "best fit" (Fit II) is fully consistent with that obtained by Gasser, Leutwyler and Sainio \cite{GLS}. This is a nontrivial result since no such constraint has been built into the procedure. 

In summary, the outcome of the present study is promising. It demonstrates that extrapolation methods based on chiral effective field theory can be successfully combined with lattice QCD results in order to bridge the gap between simulations and observables. Of course, remaining uncertainties need to be further investigated, such as corrections due to finite lattice volume and questions concerning convergence properties of the chiral expansion with quark masses exceeding 100 MeV.

Future studies will include the quark mass dependence of $r_0$, $M_N / f_\pi$ and the implicit quark mass dependence of $f_\pi$ and $g_{\pi N N}$.

\vspace{1.5cm}

We gratefully acknowledge many stimulating discussions and communications with M. Birse, M. G{\" o}ckeler, H. Leutwyler, U.-G. Mei{\ss}ner, G. Schierholz and A.W. Thomas. We thank the QCDSF-UKQCD Collaboration for providing us with their data prior to publication. 

\begin{table}[t]
 \caption{Fit results for $M_N(m_\pi)$ described in detail in the text. Fit I refers to the interpolation based on the ${\cal{O}}(p^3)$ NLO result, Eq.(\ref{massa}). Fit II and Fit III are based on the ${\cal{O}}(p^4)$ NNLO result, Eq.(\ref{massap4}), respectively with $c_3=-3.4\,{\rm GeV}^{-1}$ \cite{EM} and $c_3=-4.7\,{\rm GeV}^{-1}$ \cite{BM}.  
 \label{table1}}
 \begin{center}
  \begin{tabular}{|c|c|c|c|c|c|}\hline 
      & $M_0\;[{\rm GeV}]$ & $c_1\;[{\rm GeV}^{-1}]$ & $\hat{e}_1\;[{\rm GeV}^{-3}]$ & $A\;[{\rm GeV}^{-3}]$ & $B\;[{\rm GeV}^{-1}]$    \\ \hline \hline
 Fit I & $0.891 \pm 0.004$ & $-0.79 \pm 0.05$ & $3.5 \pm 0.6$ & - & - \\\hline
 Fit II &  $0.883\pm 0.003$ & $-0.93\pm0.04$ & - & $3.8\pm 0.6$ & $-10.4$ (fixed) \\ \hline
 Fit III &  $0.872\pm 0.003$ & $-1.11\pm0.04$ & - & $4.1 \pm 0.6$ & $-15.6$ (fixed) \\ \hline  
  \end{tabular}
 \end{center}
\end{table}

\begin{table}[t]
 \caption{The pion-nucleon sigma term deduced from the NLO and NNLO fits for $M_N(m_\pi)$ given in Table 1. 
 \label{table2}}
 \begin{center}
  \begin{tabular}{|c|c|}\hline 
      & $\sigma_N\;[{\rm MeV}]$    \\ \hline \hline
 Fit I & $43 \pm 4$  \\ \hline
 Fit II &  $49\pm 3$  \\ \hline 
  \end{tabular}
 \end{center}
\end{table}

\newpage


\end{document}